\documentclass[%
 aip,
cp,  
 amsmath,amssymb, 
reprint,%
nofootinbib
]{revtex4-2}

\usepackage{graphicx}
\usepackage{dcolumn}
\usepackage{bm}

\usepackage[utf8]{inputenc}
\usepackage[T1]{fontenc}
\usepackage{mathptmx} 

\begin{document}

\title{The ASTAROTH project}

\author{D. D'Angelo}
\email[Corresponding author: ]{davide.dangelo@mi.infn.it}
\affiliation{Universit\`a degli Studi di Milano, Dip. di Fisica, via Celoria 16, 20133 Milano, IT}
\affiliation{INFN Sezione di Milano, via Celoria 16, 20133 Milano, IT}

\author{A. Zani}
\email{andrea.zani@mi.infn.it}
\affiliation{INFN Sezione di Milano, via Celoria 16, 20133 Milano, IT}

\author{F. Alessandria}
\affiliation{INFN LASA-Milano, via fratelli Cervi 201, 20054 Segrate (Milano), IT}

\author{A. Andreani}
\affiliation{Universit\`a degli Studi di Milano, Dip. di Fisica, via Celoria 16, 20133 Milano, IT}
\affiliation{INFN Sezione di Milano, via Celoria 16, 20133 Milano, IT}

\author{A. Castoldi}
\affiliation{Politecnico di Milano, DEIB, piazza Leonardo da Vinci 32, 20133 Milano, IT}
\affiliation{INFN Sezione di Milano, via Celoria 16, 20133 Milano, IT}

\author{S. Coelli}
\affiliation{INFN Sezione di Milano, via Celoria 16, 20133 Milano, IT}

\author{D. Cortis}
\affiliation{INFN LNGS - Laboratori Nazionali del Gran Sasso, 67100 Assergi (L'Aquila), IT}

\author{G. Di Carlo}
\affiliation{INFN LNGS - Laboratori Nazionali del Gran Sasso, 67100 Assergi (L'Aquila), IT}

\author{L. Frontini}
\affiliation{Universit\`a degli Studi di Milano, Dip. di Fisica, via Celoria 16, 20133 Milano, IT}
\affiliation{INFN Sezione di Milano, via Celoria 16, 20133 Milano, IT}

\author{N. Gallice}
\affiliation{Universit\`a degli Studi di Milano, Dip. di Fisica, via Celoria 16, 20133 Milano, IT}
\affiliation{INFN Sezione di Milano, via Celoria 16, 20133 Milano, IT}

\author{C. Guazzoni}
\affiliation{Politecnico di Milano, DEIB, piazza Leonardo da Vinci 32, 20133 Milano, IT}
\affiliation{INFN Sezione di Milano, via Celoria 16, 20133 Milano, IT}

\author{V. Liberali}
\affiliation{Universit\`a degli Studi di Milano, Dip. di Fisica, via Celoria 16, 20133 Milano, IT}
\affiliation{INFN Sezione di Milano, via Celoria 16, 20133 Milano, IT}

\author{M. Monti}
\affiliation{INFN Sezione di Milano, via Celoria 16, 20133 Milano, IT}

\author{D. Orlandi}
\affiliation{INFN LNGS - Laboratori Nazionali del Gran Sasso, 67100 Assergi (L'Aquila), IT}

\author{M. Sorbi}
\affiliation{Universit\`a degli Studi di Milano, Dip. di Fisica, via Celoria 16, 20133 Milano, IT}
\affiliation{INFN LASA-Milano, via fratelli Cervi 201, 20054 Segrate (Milano), IT}

\author{A. Stabile}
\affiliation{Universit\`a degli Studi di Milano, Dip. di Fisica, via Celoria 16, 20133 Milano, IT}
\affiliation{INFN Sezione di Milano, via Celoria 16, 20133 Milano, IT}

\author {M. Statera}
\affiliation{INFN LASA-Milano, via fratelli Cervi 201, 20054 Segrate (Milano), IT}

\date{\today} 

\begin{abstract}
The most discussed topic in direct search for dark matter is arguably the verification of the DAMA claim. In fact, the observed annual modulation of the signal rate in an array of NaI(Tl) detectors can be interpreted as the awaited signature of dark matter interaction. Several experimental groups are currently engaged in the attempt to verify such a game-changing claim with the same target material. However, all present-day designs are based on a light readout via Photomultiplier Tubes, whose high noise makes it challenging to achieve a low background in the 1-6~keV energy region of the signal. Even harder it would be to break below 1~keV energy threshold, where a large fraction of the signal potentially awaits to be uncovered. ASTAROTH is an R\&D project to overcome these limitations by using Silicon Photomultipliers (SiPM) matrices to collect scintillation light from NaI(Tl). The all-active design based on cubic crystals is operating in the 87~K--150~K temperature range where SiPM noise can be even a hundred times lower with respect to PMTs. The cryostat was developed following an innovative design and is based on a copper chamber immersed in a liquid argon bath that can be instrumented as a veto detector. 
We have characterized separately the crystal and the SiPM response at low temperature and we have proceeded to the first operation of a NaI(Tl) crystal read by SiPM in cryogeny.  
\end{abstract}

\maketitle

\section{\label{sec:level1}Motivation}

The search for Dark Matter (DM) has still to provide definitively positive results, despite the continued construction of experiments with increasing target mass and exploiting different techniques and signatures. 
However, the single most important claim of DM observation so far has not yet been confirmed, nor conclusively disproved. 
For over 20~years, the DAMA experiment at the underground Gran Sasso National Laboratory (LNGS) is observing an annual modulation of the interaction rate in arrays of NaI(Tl) crystals~\cite{dama2018}. 
While several other experiments around the world fail to observe a compatible signal, their result are based on a different technique and/or target material~\cite{LZ2022, darkside2018, cresst2019}.
Achieving a next-generation detector based on NaI(Tl) with a sensitivity comparable with or higher than DAMA is very important to test of this long standing claim and, in case a dark matter induced annual modulation is present, to shed light on its nature.
In fact a few attempts are underway~\cite{anais,cosine}, although, as further discussed below, they still lack the necessary sensitivity to fully verify the DAMA signal; more programs are in preparation~\cite{sabre, picolon}. 

All current generation NaI(Tl)-based experiments share the same design of the detector modules, regardless of the exact crystal shape that can be a cylinder or an elongated cuboid with square or octagonal section. 
The crystals are always read by two 3" PhotoMultiplier Tubes (PMTs), directly coupled to the crystals, with the exception of DAMA that still uses fused silica light guides. 
The sides of the crystals are lined with a high reflectivity material to maximize light collection.

In literature NaI(Tl) emits 40-42 photons for a 1~keV electron recoil, although this number could be higher for recently produced crystals. 
With the detector design described above and a PMT quantum efficiency at best of 30-35\% in the 400-420~nm peak of Tl de-excitation, about 7-15 photoelectrons are collected per keV.
The energy region of interest (ROI) where the DAMA signal modulation is observed is 1-6~keV$_{ee}$, which means having to deal with very faint signals. 
In the ROI, the scintillation events are swamped under a ten times higher rate of noise from the PMTs that must be rejected by means of Pulse Shape Discrimination (PSD)\footnote{\label{footnote:noise}Part of this noise is related to the so called after-glow effect. Occasionally, in connection with a pulse the PMT also emits light that may cross the crystal and fire the opposite PMT. The two pulses are a few tens of ns apart and may match  the coincidence trigger condition of the detector. Other noise events have unclear origin. More details can be found for example in Ref.~\onlinecite{sabre_dry}.}.
PSD limited performance at low number of photoelectrons (ph.e.) dictates the analysis threshold, at best reaching down to 1~keV$_{ee}$.

The background contribution from PMTs is also significant and needs to be distributed over crystals of relatively large mass in the 5-12~kg range which implies growing ingots longer than 20~cm. 
The latter becomes a challenge for crystal growers when combined with the requirement of ultra-low background.

Most experiments enforce or long for a surrounding veto detector able to suppress those background components accompanied by high energy gamma emissions, such as $^{40}$K and $^{22}$Na. Such a veto could be easily accomplished by using organic liquid scintillators. These, however, impose stringent safety and environmental precautions and are eventually ruled out in particularly sensitive locations such as LNGS.

\section{Innovation}

ASTAROTH (All SensiTive ARray with lOw Threshold) aims to develop a technology that would overcome the limitations of the current generation NaI(Tl) experiments and bring this search into the necessary next step.

\begin{figure}
\includegraphics[width=.45\linewidth]{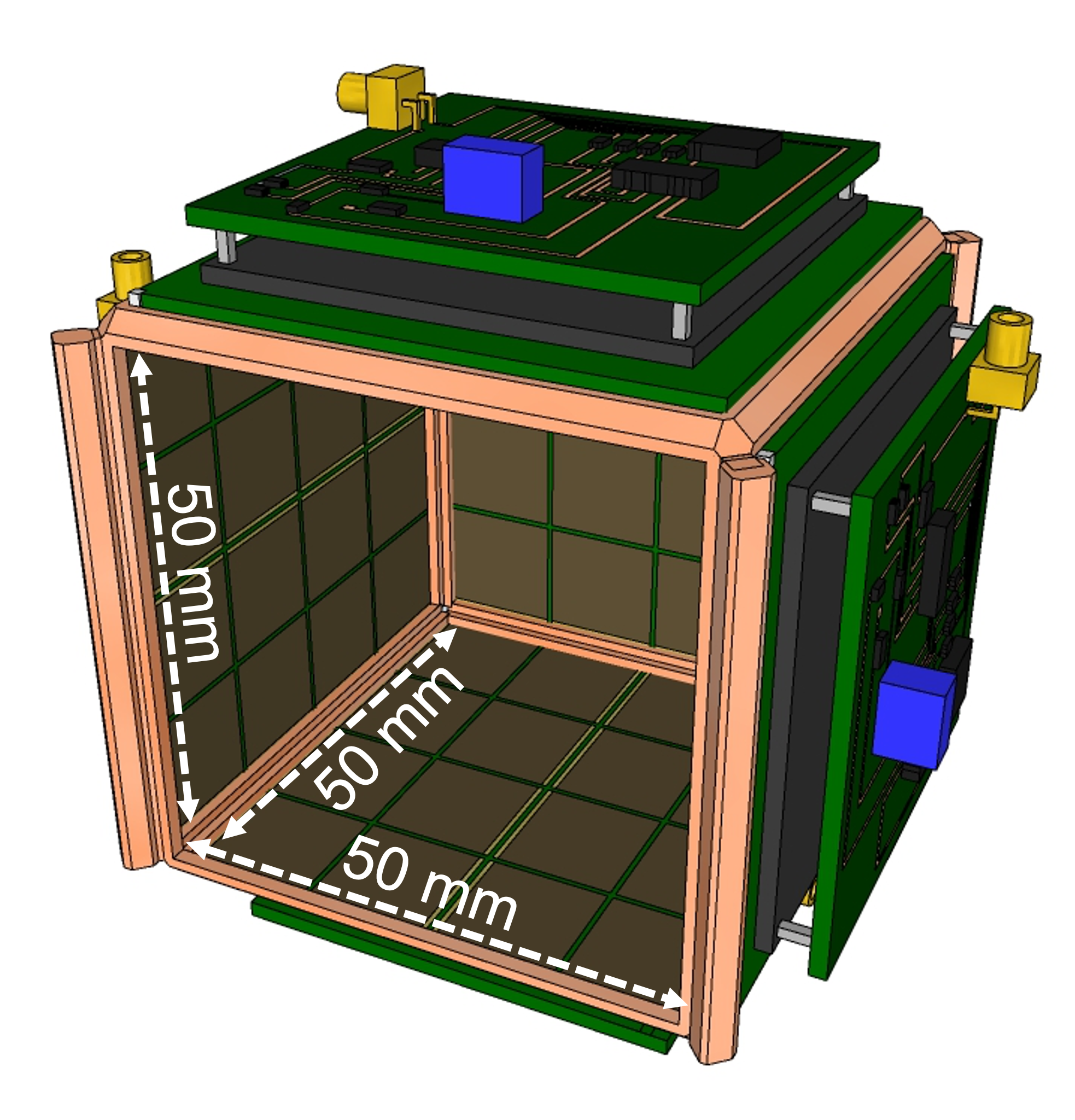}
\hfill
\includegraphics[width=.45\linewidth]{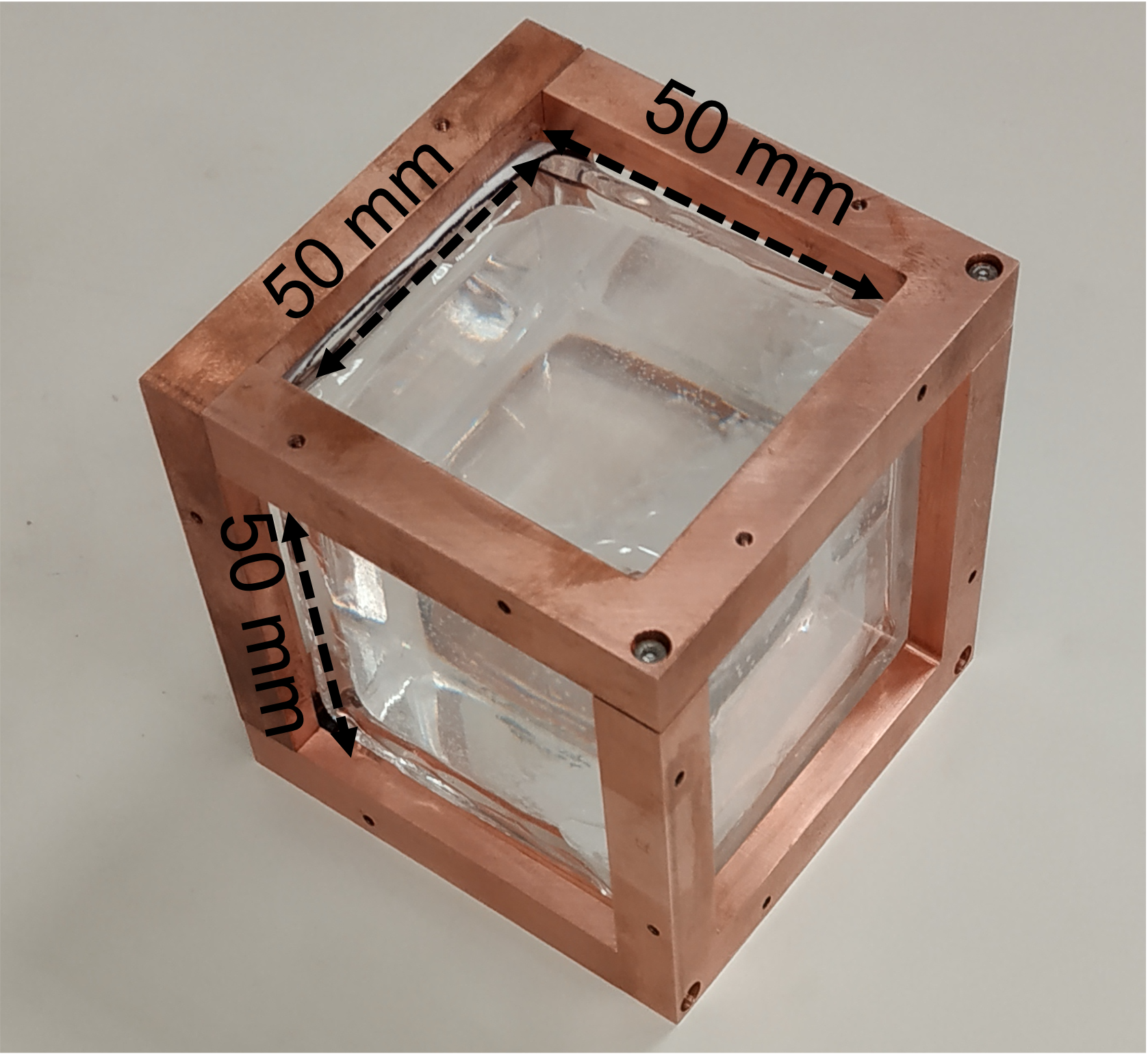}
\caption{\label{fig:sketch} (left) Conceptual 3D model of the detector module. (right) Photo of a test crystal within fused silica case and copper cage.}
\end{figure}

The all-sensitive design is based on relatively small cubic NaI(Tl) crystals with an edge of 5~cm and whose scintillation light is read by Silicon PhotoMultiplier (SiPM) arrays on all six faces (see fig.\ref{fig:sketch} left).
A future dark matter detector could scale the cubes size to 8-10~cm and still be within the reach of today's ultra-high purity crystals manufacture.

SiPMs feature multiple advantages over PMTs~\cite{acerbi}: (1) photo-detection efficiency of the recent generation NUV (Near Ultra Violet) devices is as high as 55\%, significantly outmatching PMTs; we are investigating several coupling solutions by simulation of the optical photons generation and transport and preliminary results show that over 20~ph.e.~/~keV could be reached; (2) the large population of PMT-related noise at low energy (see footnote~\ref{footnote:noise}) is not present; (3) SiPMs are essentially made of pure Si which is free from radioactive backgrounds. 

Such a detector must operate in a cryogenic environment, as the high dark count of SiPMs at room temperature makes them unsuitable for low energy applications. At liquid argon temperature, however, the dark count of SiPMs can be lower by two orders of magnitude compared to typical PMTs adopted in the field~\cite{acerbi}.
The ASTAROTH detector is based on a bath of cryogenic liquid in which a chamber hosting the crystals is immersed. 
The bath could be instrumented with additional SiPM arrays or PMTs and act as a veto detector, fully compliant with the safety and environmental regulations of any laboratory.

\section{Crystal Cooling}

ASTAROTH's innovative cryogenic system is able to cool up to two crystals coupled to SiPM arrays to an operating temperature that we plan to tune within the 87-150~K range. 
The lower bound is set by the liquid argon bath, while the upper bound is roughly the temperature above which SiPMs start to show a dark count higher than PMTs. 
The temperature dependence the crystal response (light yield and scintillation decay times) is not know a priori and is being studied in the frame of the LITE-SABRE project. 
Within the above range the optimal working temperature will be selected at run time, taking into account both the crystal and the SiPM performance.

\begin{figure}
\includegraphics[width=\linewidth]{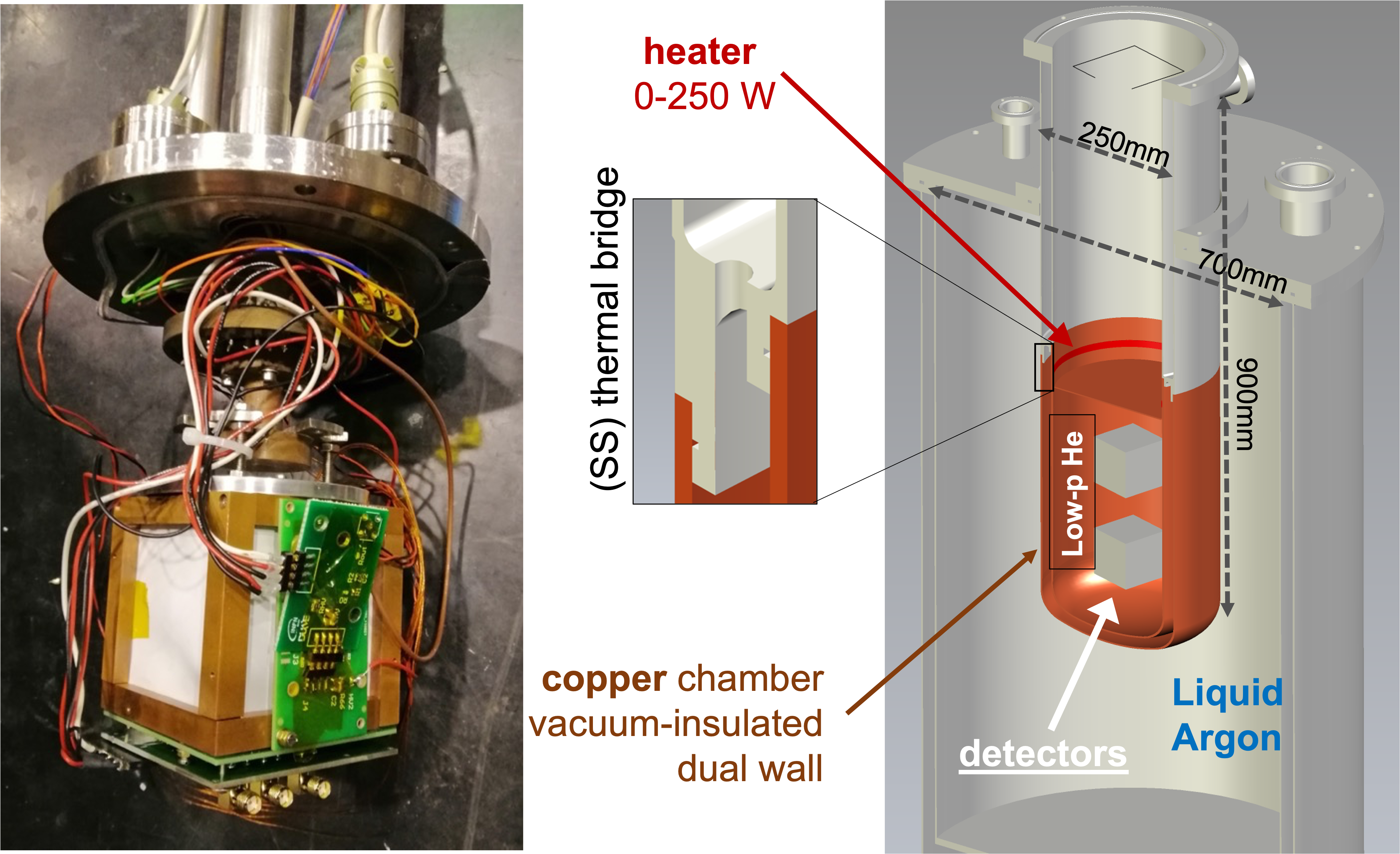}
\caption{\label{fig:cryo} (left) Detector module equipped with one SiPM array and mounted on a temporary cryostat support. (right) Cryostat CAD model with detail on the thermal bridge.}
\end{figure}

With this goal in mind we have designed a dual-walled copper chamber with a steel chimney (also dual-walled) that will be immersed in a liquid argon bath (see fig.\ref{fig:cryo}).
The copper-steel connection is brazed and vacuum will be pulled within walls.
The cooling power from the bath is transmitted to the inner volume of the chamber solely via a steel thermal bridge with a calibrated impedance.
An electric heater allows to dissipate up to 250~W of thermal power that establish a tunable dynamic equilibrium with the cooling power and stabilize the temperature at the desired value.
The inner volume will be filled with low pressure Helium gas ($\sim$~200~mbar) to provide thermal conductance toward the crystals guaranteeing at the same time slow temperature variations (preventing damage to the crystals) and uniformity at equilibrium.
Crystals will be supported by a structure hanging from the top chimney flange although the chimney volume will be separated from the chamber and segmented by means of copper/steel discs in order to limit convection cycles.

\begin{figure}
\includegraphics[width=.45\linewidth]{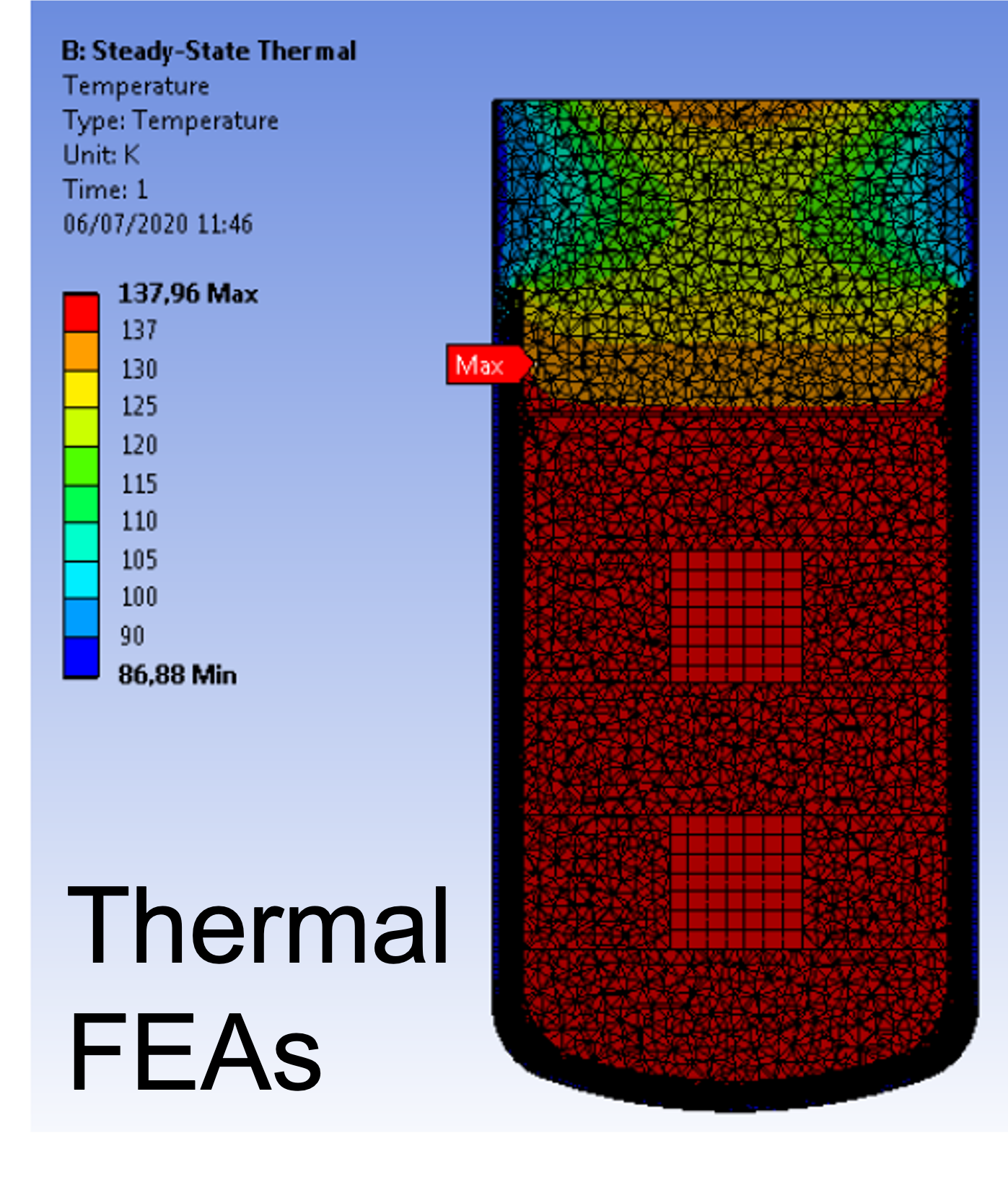}
\hfill
\includegraphics[width=.45\linewidth]{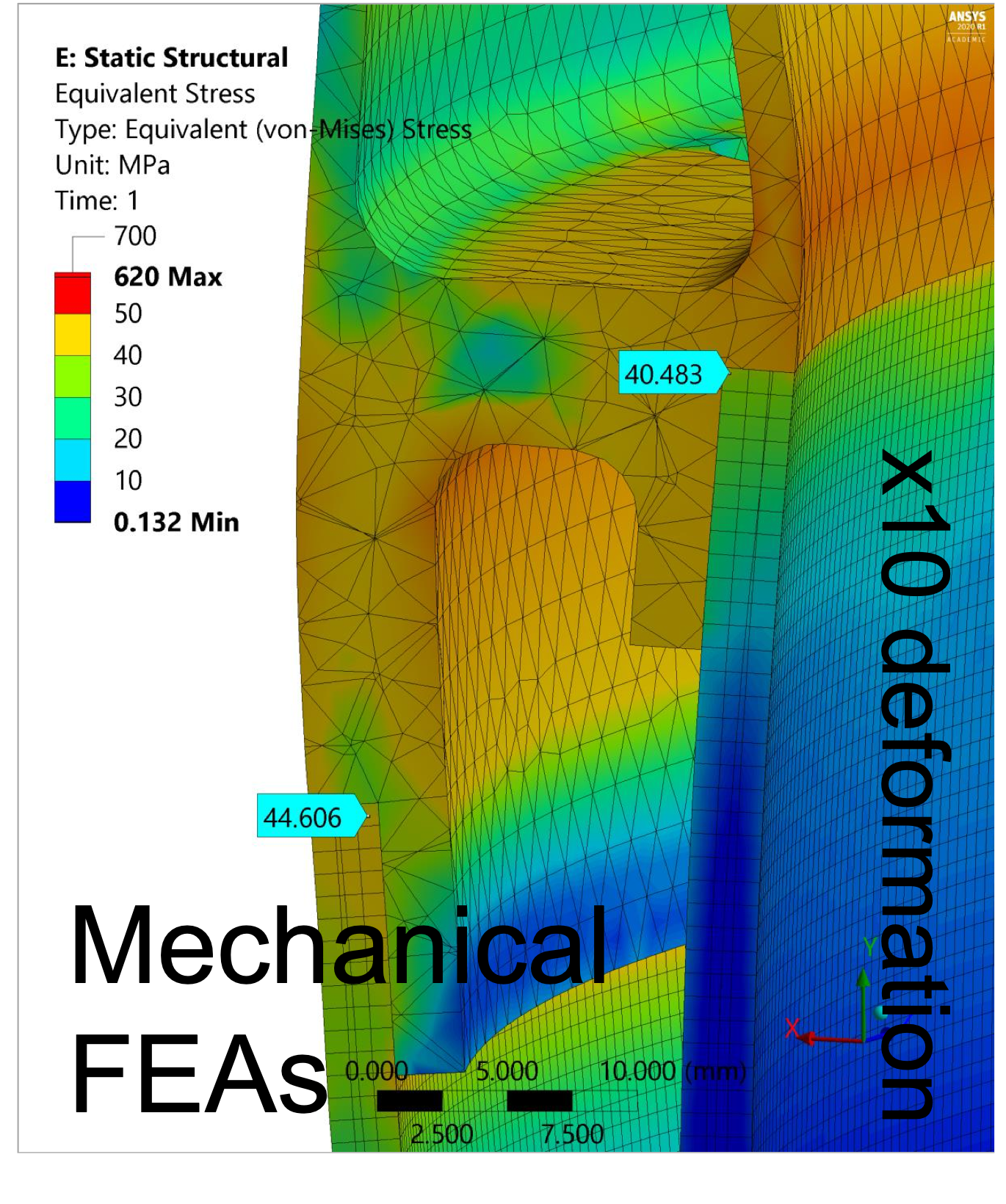}
\caption{\label{fig:fea} (left) Temperature map from the FEA thermal analysis. (right) Stress and deformation (magnified $times$~10) from the FEA mechanical analysis.}
\end{figure}

The development of this complex system has required a complete thermo-mechanical simulation of the full model before proceeding to construction (see fig.~\ref{fig:fea}). 
Using a Finite Element Analysis (FEA) method we have proceeded in two steps. 
First we have simulated the thermal map of the detector starting from the LAr cooling power and the heater dissipated power. 
After establishing the uniformity within the inner volume at the level of 0.01~K (more than suitable for dark matter search), the temperature map has been used as input for the mechanical simulation.
The goal of this second step was to verify that the maximum stress and deformations of the materials were within safety parameters, in particular in the region of the thermal bridge where a gradient of 70~K over a few mm can be reached in the most demanding operating conditions. 

The copper properties at low temperature depend on the specific composition and manufacturing process of the material. Hence, we performed multiple traction "stress \& rupture" tests at liquid nitrogen (LN$_2$) temperature on copper samples provided by the construction company, before and after reproducing the baking the copper undergoes during the brazing thermal cycle.
We also performed bending tests in LN$_2$ of brazed samples, to certify the survival of the brazed connection under the maximal thermal deformations foreseen by the mechanical simulation.

The cryostat was delivered in June 2022 and will be commissioned soon at LASA (Laboratorio Acceleratori e Superconduttività Applicata) facility near Milan.

\section{R\&D comparing technical design solutions}

ASTAROTH will compare several technical solutions regarding the crystal encasing, the SiPM technology and layout and the front-end electronics.

NaI is highly hygroscopic and we are developing a gas-tight case that allows easy handling and mounting of the detectors, while preserving transparency on all six faces. 
This is a challenging task, considering that the tightness must be kept over repeated thermal cycles to liquid argon temperature and that low radioactivity materials are desirable. 
The baseline option is a fused silica case that we are co-developing with Hilger Crystals (UK) company, where five sides are soldered (with a special quartz-on-quartz technique) and the lid is glued after the crystal's insertion. 
The alternative solution is to use acrylic and we have prepared a prototype with the help of the Alberta University group (experts on acrylic for DEAP-3600~\cite{deap} detector).
Cases are placed inside a copper cage that allows both mounting the SiPM arrays and hanging the detectors in the chamber (see fig.\ref{fig:sketch} right). 

\begin{figure}
\includegraphics[width=.30\linewidth]{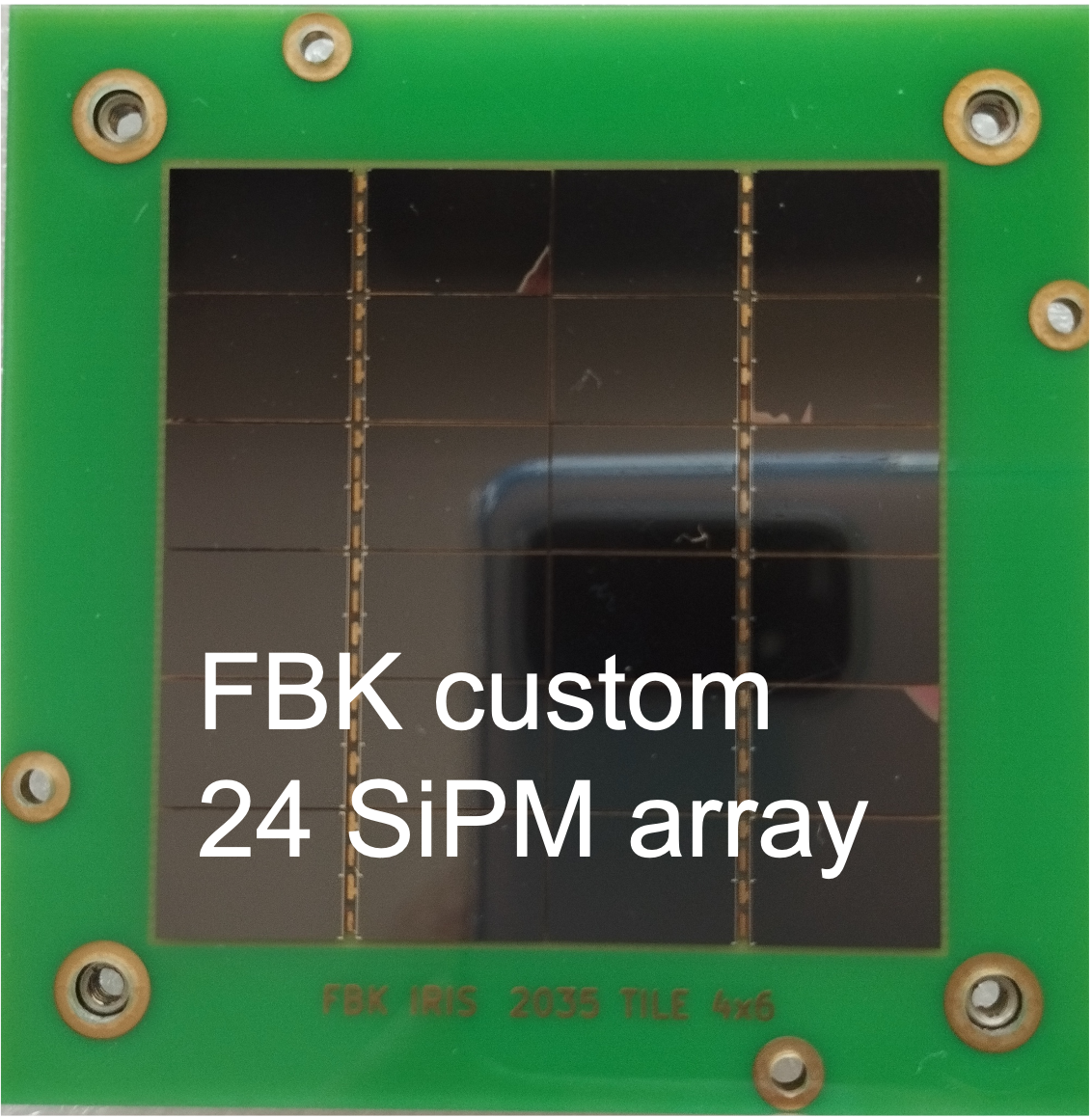}
\hfill
\includegraphics[width=.30\linewidth]{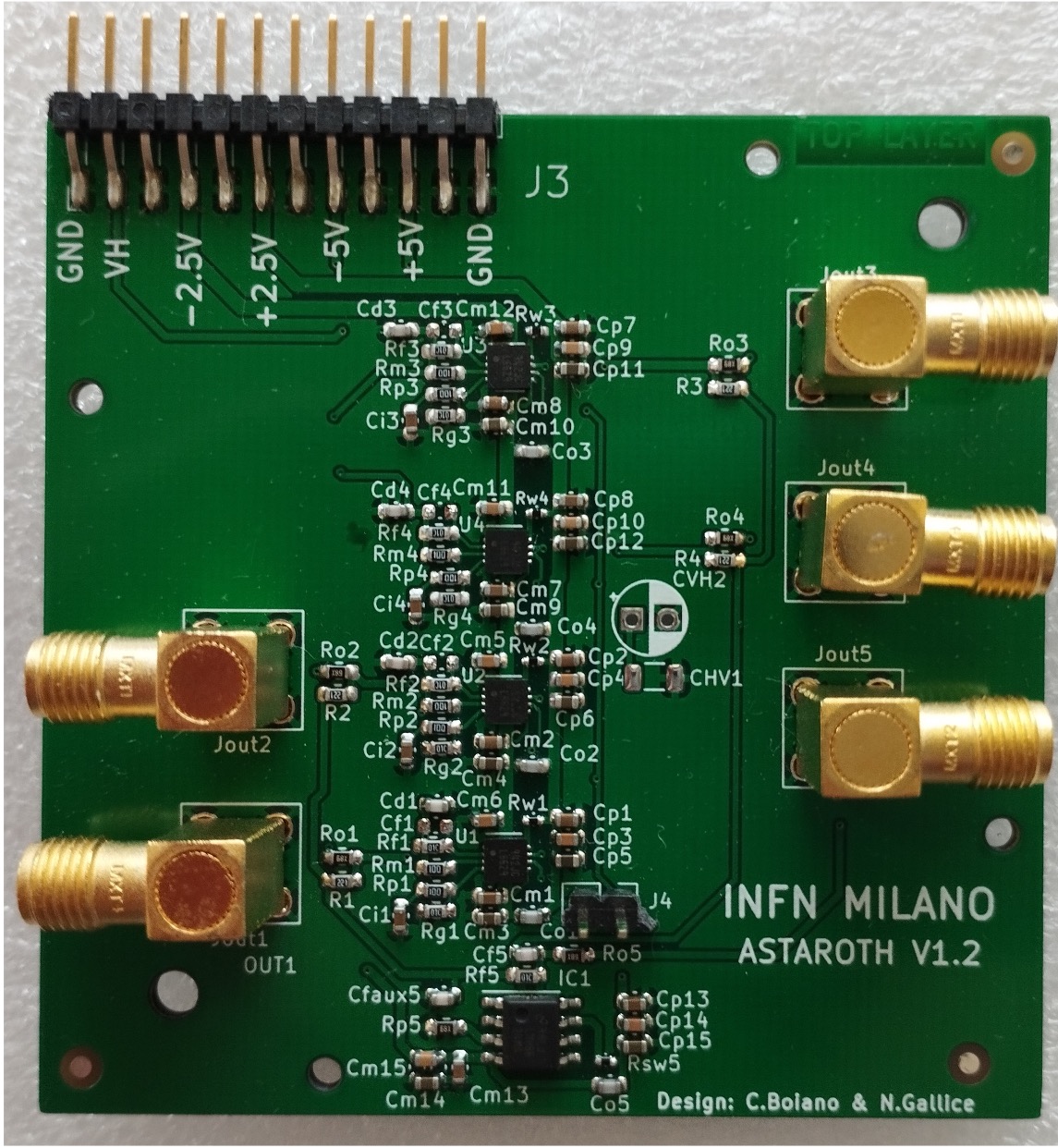}
\hfill
\includegraphics[width=.30\linewidth]{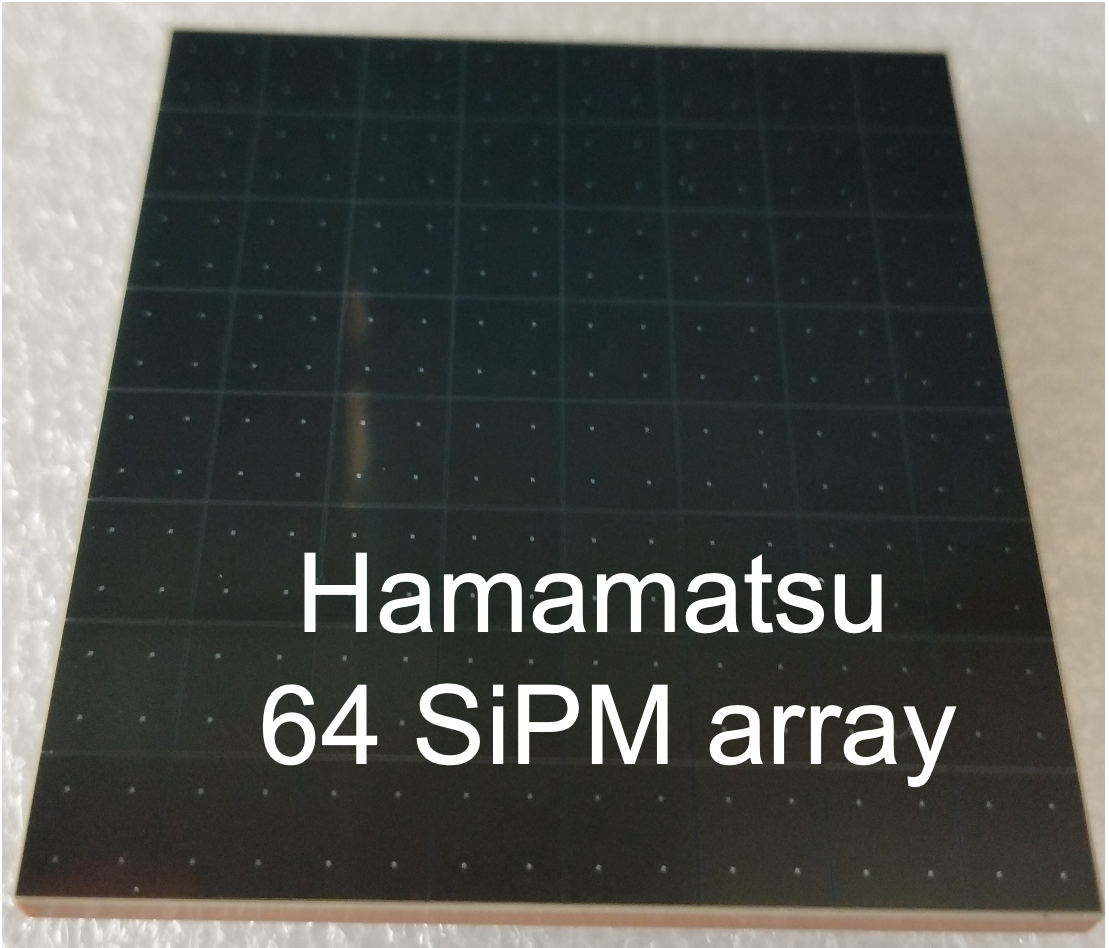}
\caption{\label{fig:sipm} (left) Custom FBK 24-devices array. (center) Croygenic amplifier designed for FBK array. (right) Hamamatsu commercial 64-devices array. All sensitive areas are 50$\times$50~mm$^2$.}
\end{figure}

Concerning SiPMs, we work with two vendors that have NUV devices suitable for cryogenic temperatures, namely Fondazione Bruno Kessler (FBK) and Hamamatsu Photonics (HPK). 
With FBK we have developed two 50$\times$50~mm$^2$ arrays based on their NUV-HD-Cryo devices \cite{nuv}. 
These are high-density, low field SiPM with a 40~\textmu m microcell size, wire bonding and epoxy coating.
The first array (fig.~\ref{fig:sipm} left) features 24 devices as big as 8$\times$12~mm$^2$ with on-board ganging in bunches of six following a scheme "2s3p", where two devices are in series and a parallel of three such pairs is read as a single channel.
The cryogenic front-end trans-impedance amplifier is a coplanar 50$\times$50~mm$^2$ board with four channels plus an additional sum stage (fig.~\ref{fig:sipm} center) and is loosely based on the scheme of Ref.~\onlinecite{dincecco}.
HPK provided us with an array of 64 devices, 6$\times$6~mm$^2$, of the series S13361-6050 (fig.~\ref{fig:sipm} right). 
These have 50~\textmu m microcell size, Through-Silicon-Via technology (that allows them to be mounted closer together with respect to wire bonded devices), silicon resin coating and no on-board ganging. 
For this array the front-end electronics is under development with a broad flexibility to read different channel multiplicities (1, 4, 16, or 64).
The second array of FBK is being manufactured with NUV-HD-Cryo devices but following the same design of the HPK array. 
In fact even connectors are identical and the two could be equipped with the same electronics.

\section{Technological Outlook}

In parallel to the development of the amplifiers based on discrete electronics, we are also aiming to an integrated circuit solution. 
The baseline is a board designed for DarkSide-20k~\cite{darkside20k} outer detector (outdated in the new design) designed by INFN Genova and mounting an ASIC developed by INFN Torino using LFoundry 110~nm technology. 
The chip is a fully analog summing amplifier that can read the first of the FBK arrays with four channels.
We plan to build on this solution by adding the digitization on-chip of the sum waveform as well as single-device charge digital readout. This would make a redundant energy estimator and also add valuable information that can help reject localized surface background events. For example $^{210}$Pb implanted in the surfaces is reported as dominant by several experiments.
The characterization of the technology at cryogenic temperatures for digital applications has never been done and has a self-standing interest for the integrated circuits community. 
To this aim we have designed a test chip which LFoundry has recently delivered and we will test it in the ASTAROTH cryostat.

As SiPM are essentially free from radioactive components, the background contribution from the light sensors is actually dictated by the front-end electronic, namely Printed Circuit Board (PCB), components and connectors.
Keeping this in mind, our final goal is to mount the SiPMs and the ASIC on the two faces of a single PCB (no connectors), using a low radioactivity support such as Arlon (used in DarkSide-20k) or Pyralux (used in SABRE) to produce a compact, low noise and low radioactivity light sensors that could replace PMTs for a variety of cryogenic applications in Astroparticle Physics.

\begin{acknowledgments}
ASTAROTH is financed by INFN under CSNV. LITE-SABRE Project (P.I. Dr. A. Zani) has received funding from the European Union’s Horizon 2020 research and innovation programme under the Marie Skłodowska-Curie grant agreement No 754496.
\end{acknowledgments}

\nocite{*}
\bibliography{bibliography}

\end{document}